\pdfoutput=1

\documentclass[a4paper,amsmath,amssymb,twocolumn]{revtex4} 

\usepackage{graphicx} \input{epsf}

\newcommand{\tdot}[1]{{\hskip2pt\ddot{\null}\hskip2.5pt
\dot{\null}\kern -5pt {#1}}}

\begin{document}

\title{Spatiotemporal Synchronization of Drift Waves in a Magnetron Sputtering Plasma}

\author{E. Martines$^{1}$, M. Zuin$^{1}$, R. Cavazzana$^{1}$, J. Ad\'{a}mek$^{2}$, V. Antoni$^{1}$, G. Serianni$^{1}$, M. Spolaore$^{1}$, N. Vianello$^{1}$}
\affiliation{$^{1}$Consorzio RFX, Associazione EURATOM-ENEA sulla Fusione, Padova, Italy
\\$^{2}$Institute of Plasma Physics AS CR, Prague, Czech Republic}

\date{\today}
\begin{abstract}
A feedforward  scheme is applied for drift waves control in a magnetized magnetron sputtering plasma.
A system of driven electrodes collecting electron current in a limited region of the explored plasma is used to
interact with unstable drift waves. Drift waves actually appear as electrostatic modes
characterized by discrete wavelengths of the order of few centimeters and frequencies
of about 100 kHz.
The effect of external quasi-periodic, both in time and space, travelling perturbations is studied.
Particular emphasis is given to
the role played by the phase relation between the natural and the imposed fluctuations.
It is observed that it is possible by means of localized electrodes, collecting currents
which are negligible with respect to those flowing in the plasma, to transfer energy to one single mode
and to reduce that associated to the others.
Due to the weakness of the external action, only partial control has been achieved.
\end{abstract}
\maketitle

\section{Introduction}
Active control of plasma fluctuations is a line of research which
has potentially important implications for controlled thermonuclear fusion activities.
Active magnetic coils have been, indeed, widely and successfully used in fusion devices for MHD control, and an electrostatic approach has been recently proposed with the aim of inducing given harmonics on the global equilibrium by means of distributed electrodes at the edge plasma \cite{Nebel}.
Moreover, at the plasma edge, electrostatic fluctuations of various nature are believed to be responsible for cross-field energy and particle
transport processes in fusion devices. For
example, it has been shown in a linear machine that by driving one of the drift
wave modes present in the plasma the background turbulence level can be reduced
\cite{Schroder}. More recently, also the nonlinear interaction of drift waves
with externally driven currents
has  been investigated \cite{Brandt}.
The synchronization of dust acoustic waves in a dusty plasma has been also successfully achieved using a negatively biased driving electrode \cite{Suranga}. In the Large Plasma Device (LAPD) the gradient-driven instability has been suppressed by the beating of two independent externally driven shear-Alfvén waves \cite{Auerbach}.

In this paper the first attempt of synchronizing naturally
occurring electrostatic fluctuations to an externally applied resonant pattern
in a low temperature magnetron plasma is described.
Magnetrons are plasma sputtering devices widely used for thin film deposition
applications. They can be broadly divided into two categories, cylindrical and
planar magnetrons. In both types, a glow discharge is established between a
cathode and an anode in the presence of a magnetic field confining the electrons
in a plasma region with a relatively high density and ionization rate.
A high flux of ions is thus created by ionization of the working gas and
accelerated towards the cathode to induce the sputtering of a
target placed over it.

A range of electrostatic fluctuations has been
observed in cylindrical and planar magnetron plasmas, powered in DC \cite{Thornton,Sheridan,Kudrna},
 in a pulsed way \cite{Ehiasarian} and at radiofrequency \cite{Wu}.
The main aim of studying fluctuation phenomena in such plasmas
is to explain the cross-field transport rate and its role in the efficiency of the
plasma processes involved. As said above, this would have an interest
not only for the applicative perspective of
improving the deposition rate, but also for plasma physics in general. In fusion plasmas, for example,
fluctuations driven by a wide kind of unstable modes are believed
 to be at the origin of confinement
degradation \cite{Hidalgo}.

This paper is devoted to the description of the results of an experimental activity aimed at synchronizing
spontaneous electrostatic instabilities by means of localized electrodes immersed in a magnetron plasma.
The paper is organized as follows. Section II is devoted to the description of the experimental setup, section III to the
description and the discussion if the experimental results. Conclusion are then presented. In the Appendix the deduction of the linear dispersion relation for the instability under investigation is given.

\section{Experimental Setup}
\begin{figure}
\includegraphics[width=1.\columnwidth]{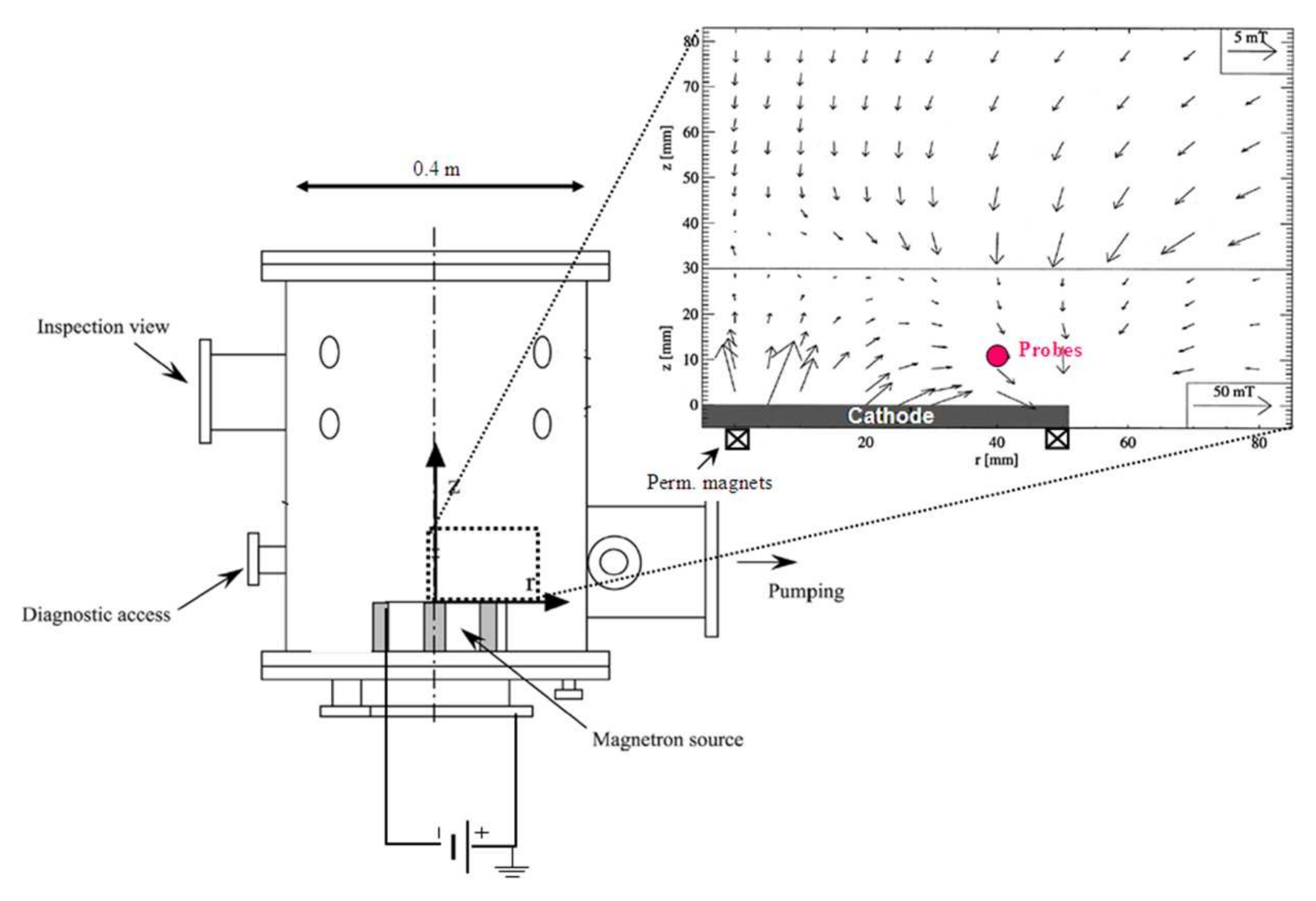}
\caption{Scheme of the magnetron sputtering device, with a detail of the
plasma region under investigation showing the map of the magnetic field lines
and the location of the probes.}
\label{schema_magnetron}
\end{figure}

The experimental activity here described has been performed on
a planar magnetron sputtering device, whose layout is shown in Fig. \ref{schema_magnetron}, constituted by a
stainless steel vacuum chamber (500 mm
height and 400 mm diameter) and by
a sputtering source (the cathode), 10 cm diameter, placed at the bottom of
the vacuum chamber itself; $z$ is the vertical source axis, with its origin on
the target upper surface and directed upward, and $r$ the radial coordinate.
The device has been operated with a DC power supply and Argon as filling
gas. The magnetic field above the source is provided by 24 pairs of
permanent magnets circularly equally spaced at about 50 mm from the axis, plus a further
pair placed in the center. Lateral magnets have a magnetic polarization opposite to that
of the central pair.
In Fig. \ref{schema_magnetron} a detail of the location of the magnets and of the magnetic field map above the magnetron source is also shown. The magnets
are placed under the target in a water cooled copper bulk, and an iron plate
is placed under them.

\begin{figure}
\includegraphics[width=0.8\columnwidth]{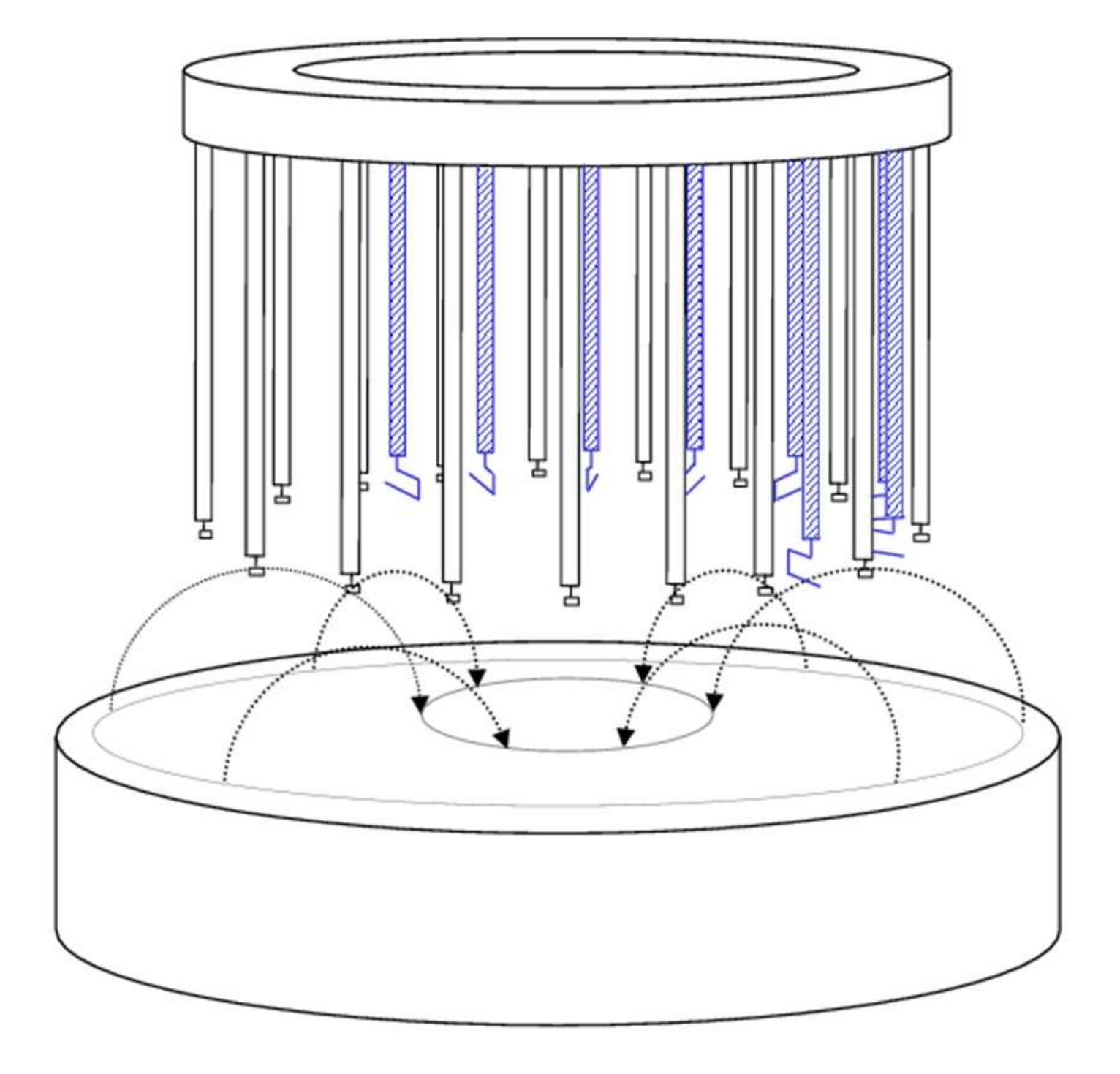}
\caption{Scheme of the diagnostic and of the active control electrodes apparatus disposed above the
cathode surface at $z = 13$ mm. The electrodes can be recognized for their
U-shape. Dashed curves
represent the magnetic field lines intersecting the plasma region under investigation.}
\label{Magnetron_probes_system}
\end{figure}

In order to investigate the spatial and temporal
plasma fluctuation dynamics a diagnostic tool had been developed, as described in Ref. \cite{Martines04}. This
diagnostic system consists of 16 equally spaced Langmuir probes forming a
circular array with a radius of 40 mm.
Each probe electrode is a stainless steel cylinder, 1.5
mm in radius and 1 mm long, connected to a stainless steel wire with 0.8 mm
diameter, for a total collecting area of about 16 mm$^2$. The wire is housed in
a quartz tube, having inner and outer radii of 1 mm and 3 mm and a length of
120 mm. The probes are kept in place by means of an aluminum disc, which acts
as supporting structure. As shown in Fig. \ref{Magnetron_probes_system}, the
system was placed coaxially with the cylindrical vacuum chamber, with the probes
located on a plane parallel to the cathode surface at $z = 13$ mm, such position being indicated
also in Fig. \ref{schema_magnetron}.
The position was chosen consistently with the diagnostics described  in Refs. \cite{Martines01,Martines04}.

\begin{figure}
\includegraphics[width=0.8\columnwidth]{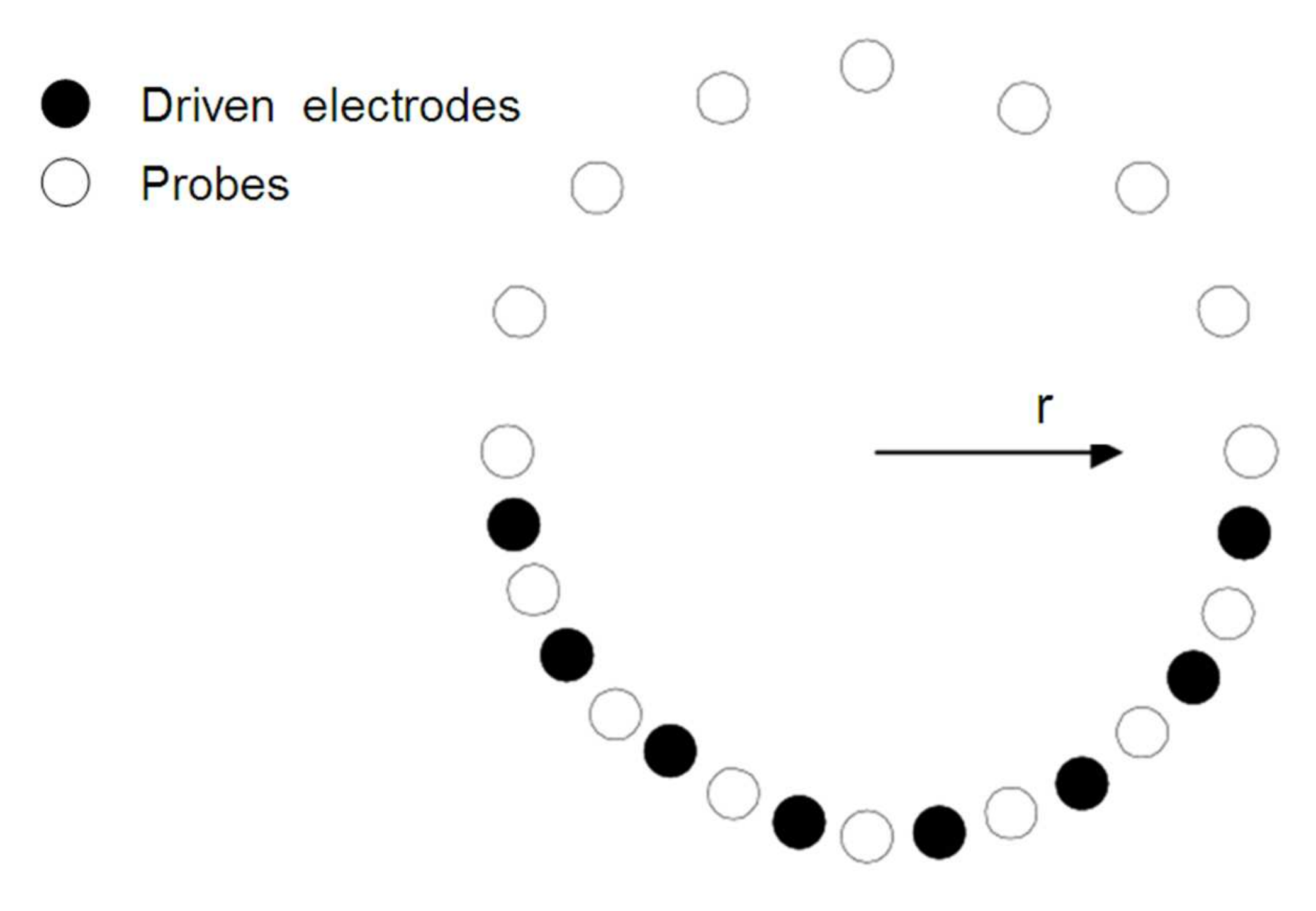}
\caption{Probes and driven electrodes distribution, as seen from above.}
\label{Magnetron_probes_electrodes}
\end{figure}

The active control system consists of 8 electrodes placed at the same $z$ and
radial position of the probes. The electrodes are $U-shaped$ stainless steel
cylinders (see Fig. \ref{Magnetron_probes_system}), 1 mm diameter, 35 mm long,
for a total collecting area of 110 mm$^2$, which is almost one order of
magnitude larger than the probe area. As shown in Fig.
\ref{Magnetron_probes_electrodes}, the 8 electrodes used as drivers are
azimuthally displaced  so that each electrode lays between two adjacent probes.

It is important to note that this layout has been chosen to obtain a drive
system which is local, affecting only half of the domain. This is an important
condition, which differentiates our control experiments from previous
ones, where almost the whole plasma volume was affected by equally
distributed active electrodes \cite{Schroder,Brandt}.
\

The fluctuation measurements are collected by connecting the probes to ground through a resistance chosen
to be 1 k$\Omega$ , and measuring the potential fall $\delta$V on the resistance itself. Such
configuration, which puts our measurement in between the real floating potential value and the negative
potential for the ion saturation current on the Langmuir characteristics, must be considered as a
compromise to obtain a larger frequency bandwidth (usually attainable by measuring the ion saturation
current fluctuations) without the use of high voltage power supplies for such a large number of probes. Due
to the closeness of such measurement to the floating potential value, we will hereafter name it $V_f$. The
data were collected by means of a digital oscilloscope with isolated inputs, sampled at 2 MHz with 12-bit
resolution using the oscilloscope itself. Records of 200 ksamples were captured. The chosen values for the
two main control parameters, namely the discharge current and the neutral gas pressure are $I = 0.6$ A and
$P = 1$ Pa, respectively, corresponding to a discharge power of 260 W.

\begin{figure}
\includegraphics[width=1.\columnwidth]{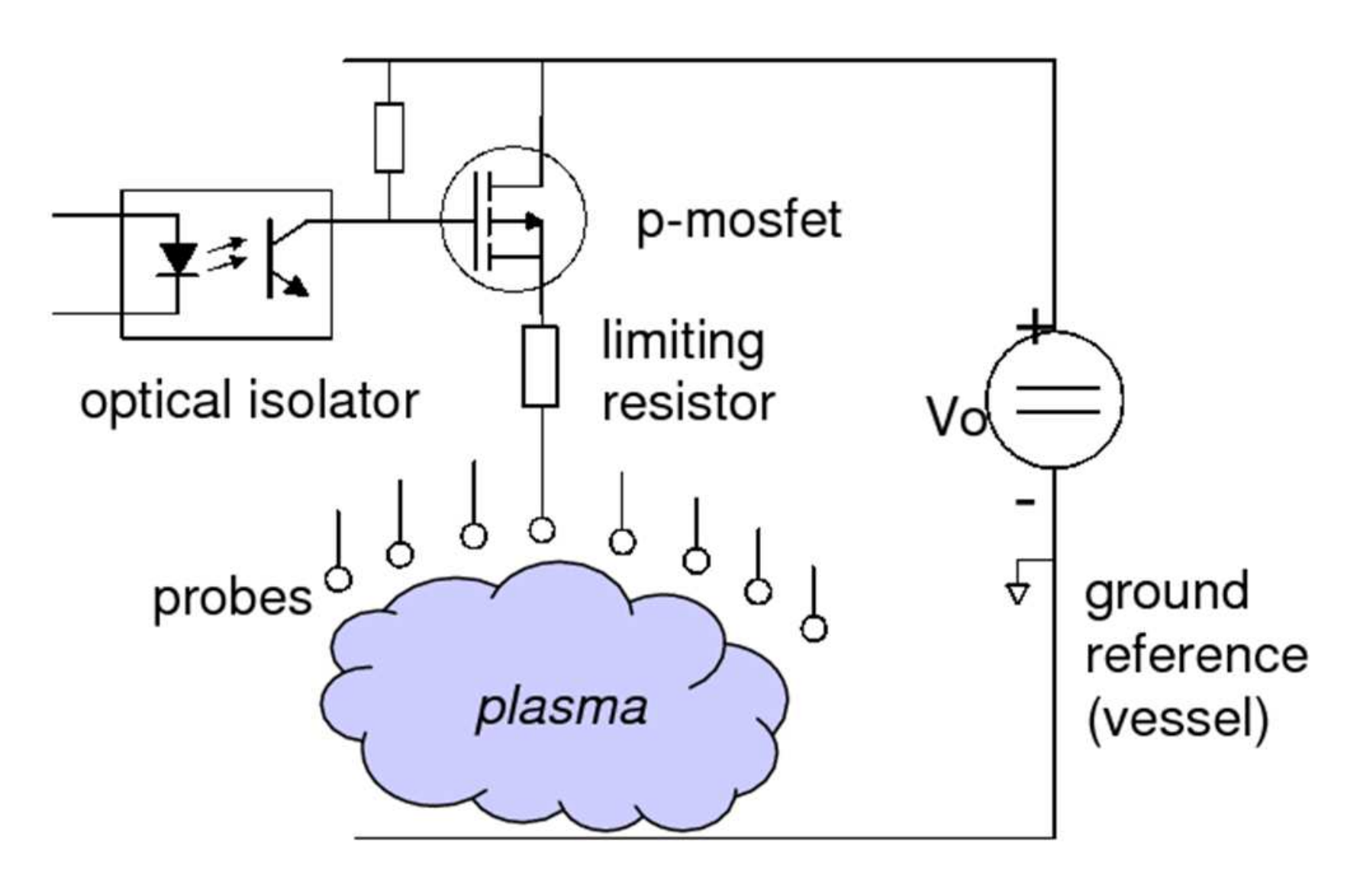}
\caption{Scheme of principle of the electrodes driving system.}
\label{Magnetron_drive_scheme}
\end{figure}

Regarding the control system, the scheme adopted in these experiments allows
to create in the 8 drivers a travelling pattern with a defined azimuthal
periodicity at a chosen frequency. This is performed, in practice, by connecting
in sequence each electrode to a power supply delivering a constant voltage
$V_{d}$ larger than the floating potential using a Mosfet. The Mosfets are
controlled by a programmable function generator, so as to create the correct
spatiotemporal pattern. In this way, a train of positive (i.e. from the probe
to the plasma) current peaks is created, forming a perturbation with the
appropriate wavenumber and frequency. The wave is clearly non-sinusoidal, and
indeed this approach was chosen having recognized that only driving the probe
to voltages higher than the floating potential, towards the electron part of
the $I-V$ characteristic, yields a substantial current. A scheme of principle of
the driving system is reported in Fig. \ref{Magnetron_drive_scheme}.

It is important to observe here that, due to the rather small collecting
surfaces of the probes, only small currents (of the order of few mA) can be
drawn, so that the applied perturbation is weak ($\leq 1\%$) when compared to the global
discharge properties.

\section{Experimental Results}

\begin{figure}
\includegraphics[width=1.\columnwidth]{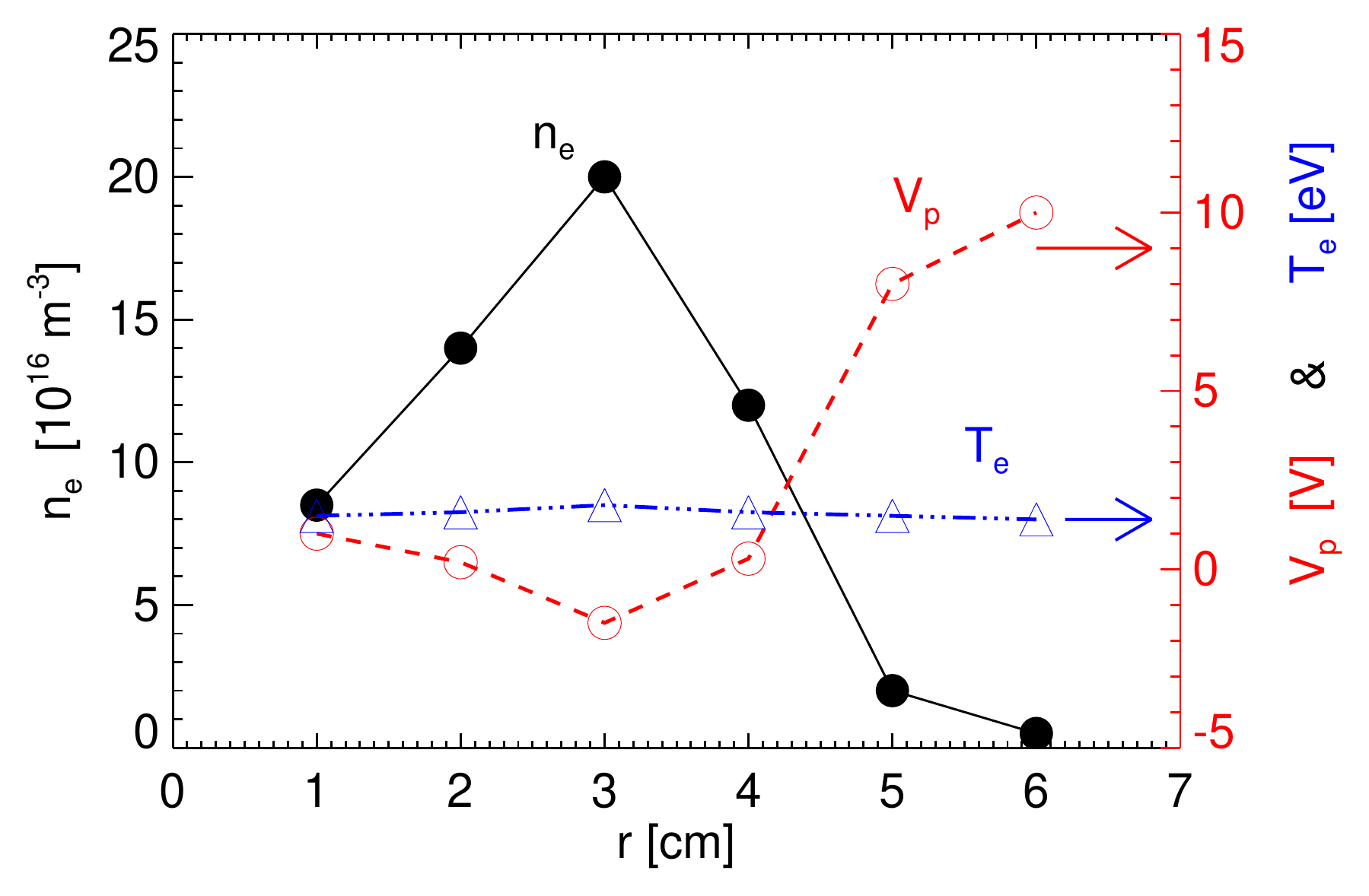}
\caption{Radial profiles of electron density $n_e$ (full dots), temperature $T_e$ (circles) and plasma potential $V_p$ (triangles)
measured at $z=13$ mm from
the cathode surface. The two latter quantities refer to the righthand side y-axis. }
\label{Magnetron_mean_profiles}
\end{figure}

In Fig. \ref{Magnetron_mean_profiles} the horizontal profiles of the time averaged plasma
parameters at $z=13$ mm, i.e. in the position of the probes and the driven electrodes, namely plasma
potential $V_p$, electron density $n_e$ and temperature $T_e$,
as measured and described in \cite{Martines04}, are shown. These data
are given for completeness. Indeed,  they are a key ingredient
in the theoretical interpretation of the fluctuation phenomena we are here trying to control in terms of
drift waves \cite{Martines01,Martines04}.

The density radial profile is found to be peaked at $r\sim30$ mm, inside the magnetic trap, where
the plasma potential has a minimum, which is
different from what obtained
 in other magnetron devices  \cite{Bradley01}. The electron
temperature inside the trap is in the range 1.4 - 2 eV, without any clear
dependence on the radial coordinate (ions are at room temperature). It is worth to note that, as typical of this kind of plasmas,
the electrons have been observed to be characterized not by a simple Maxwellian distribution, but by the
superposition of two main populations at different energies. The colder component, the one we are
referring to in this paper,
 is the dominant one in terms
of particle number.  The hotter component of the electron distribution has an
electron temperature of about 15 eV at a density that is about three orders
of magnitude lower than the colder one \cite{Serianni}.

\begin{figure}
\includegraphics[width=1.\columnwidth]{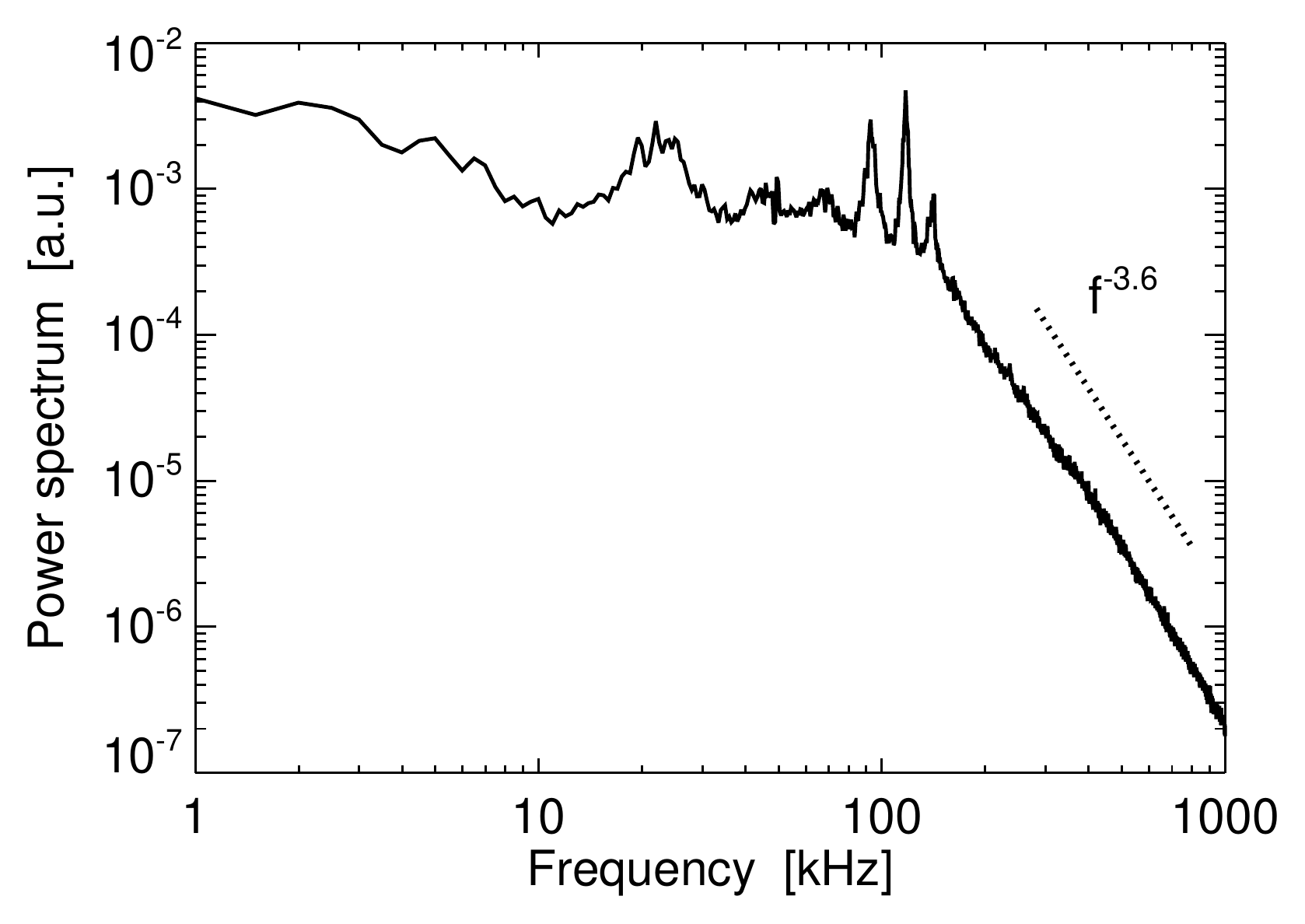}
\caption{Power spectrum deduced from a single $V_f$ probe signal. A power law interpolation ($f^{-\alpha}$,
with $\alpha \simeq 3.6$)
of the high
frequency part of the spectrum is proposed as a dashed line.}
\label{Sf_esempio}
\end{figure}

Turning now to the fluctuation measurements, we report here (see Fig. \ref{Sf_esempio})
an example of the spectra obtained from one of the 16 probes composing the azimuthal array.
The spectrum is found to be characterized by the presence of few coherent peaks in the frequency range
between 10 and 200 kHz, more pronounced above 80 kHz. The higher frequency part of the spectrum
exhibits a power law decay, $f^{-\alpha}$, with the spectral index $\alpha \simeq 3.6$.

\begin{figure}[!t]
\includegraphics[width=1.\columnwidth]{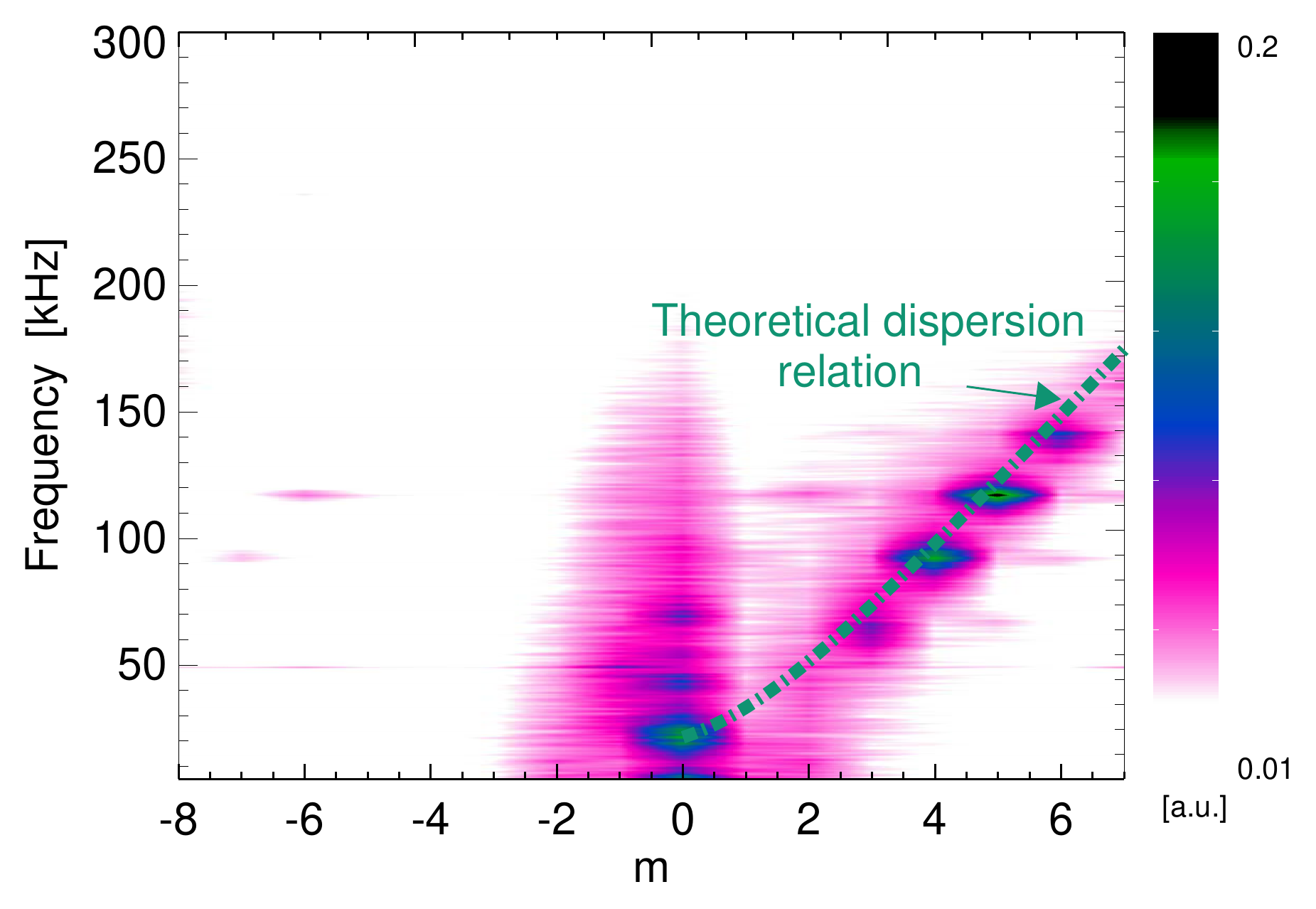}
\caption{Contour plot of the frequency and wavenumber resolved spectrum $S(m,f)$  of the fluctuation.
The superimposed dashed curve represents the dispersion relation given by
eq. \ref{Magnetron_dispersion_relation}.}
\label{Magnetron_dispersion_exp_th}
\end{figure}

By means of the full array of 16 probes the dispersion relation (i.e. the relation between the frequency
peaks and the corresponding wave numbers) along the azimuthal direction can be deduced. In Fig. \ref{Magnetron_dispersion_exp_th}
the $S(k,f)$ spectrum
is shown in a color-coded contour, the color being associated (in a.u.) to the spectral density.
The azimuthal component of the wavevector $k_y$ is actually expressed in terms of
the mode number $m=k_y r$, being $r$ the radius at which
the measurements are taken.
It is evident that the three dominant frequency
peaks, at $f>60$ kHz, correspond to different azimuthal Fourier modes, with $m=3,4,5$ ($m = 6$ and $m = 7$ are also slightly visible).
Along with the part of spectrum propagating in the electron diamagnetic direction an important
$m=0$ component is present, exhibiting a broadband activity superposed to a peak located at around 20 kHz and
to its higher harmonics.

The overplotted curve is the
theoretical linear dispersion relation:
\begin{equation}\begin{split}
\omega^2 +
\omega[\frac{\Omega_i}{\nu_i}(k_x-\frac{\Omega_i}{\nu_i}k_y)v_E+i(\nu_i-Z)]-
\\
-k^2C_s^2+Z\nu_i-i\Omega_i[(\frac{\Omega_i}{\nu_i}k_y-k_x)v_E+k_xv_{de}]=0
\label{Magnetron_dispersion_relation}
\end{split}\end{equation}
whose deduction, exposed in
refs. \cite{Martines01,Martines04}, is briefly resumed in the Appendix.
 Here $C_s$ is the ion sound velocity, $C_s\simeq(T_e/m)^{1/2}$, $v_E$ is the
$\textbf{E}\times\textbf{B}$ velocity, $\nu_i$ is the collision frequency of
ions with neutrals, $\Omega_i$ is the ion cyclotron frequency multiplied by
$2\pi$, k is the wavevector and $v_{de}$ is the electron diamagnetic drift
velocity $v_{de}=-T_e/(enB)(dn/dx)$).

The real part of the root of the theoretical dispersion relation, i.e the solution
$f=f(k_y)$, being $f=\omega /2\pi$, of this equation is found to be in
good agreement with the experimental peaks for the case considered. Note that the $k_x$
and $k_z$ values are taken as free parameters in a nonlinear fit procedure, and that the
resulting wavelengths are compatible with the geometry of the chamber and of the magnetic field lines
($\lambda_x\approx4$ cm and $\lambda_z\approx20$ cm).
A detailed characterization of the properties
of the fluctuations in a wide range of neutral gas pressure has been performed and described in Ref. \cite{Martines04},
along
with a comparison between the linear theoretical growth rate and the observed mode amplitude.
The agreement between the theoretical and the experimental dispersion relation allows to  interpret the fluctuations under study in terms of a $\textbf{E}
\times\textbf{B}/$density gradient instability, also known as \textit{neutral
drag} instability.

It must be said that peaks at frequencies of the order of few kHz in the fluctuation power spectrum have been already
observed in a DC cylindrical magnetron discharge by Kudrna et al.
\cite{Kudrna}, and at the ion cyclotron frequency
by Sheridan and Goree in a
planar magnetron system \cite{Sheridan}.
It is worth to add that the observed instabilities exhibit almost the same properties
of those characterizing the plasma of a high power impulse magnetron sputtering, as recently reported \cite{Ehiasarian}.

\begin{figure}
\centering
\includegraphics[width=1.\columnwidth]{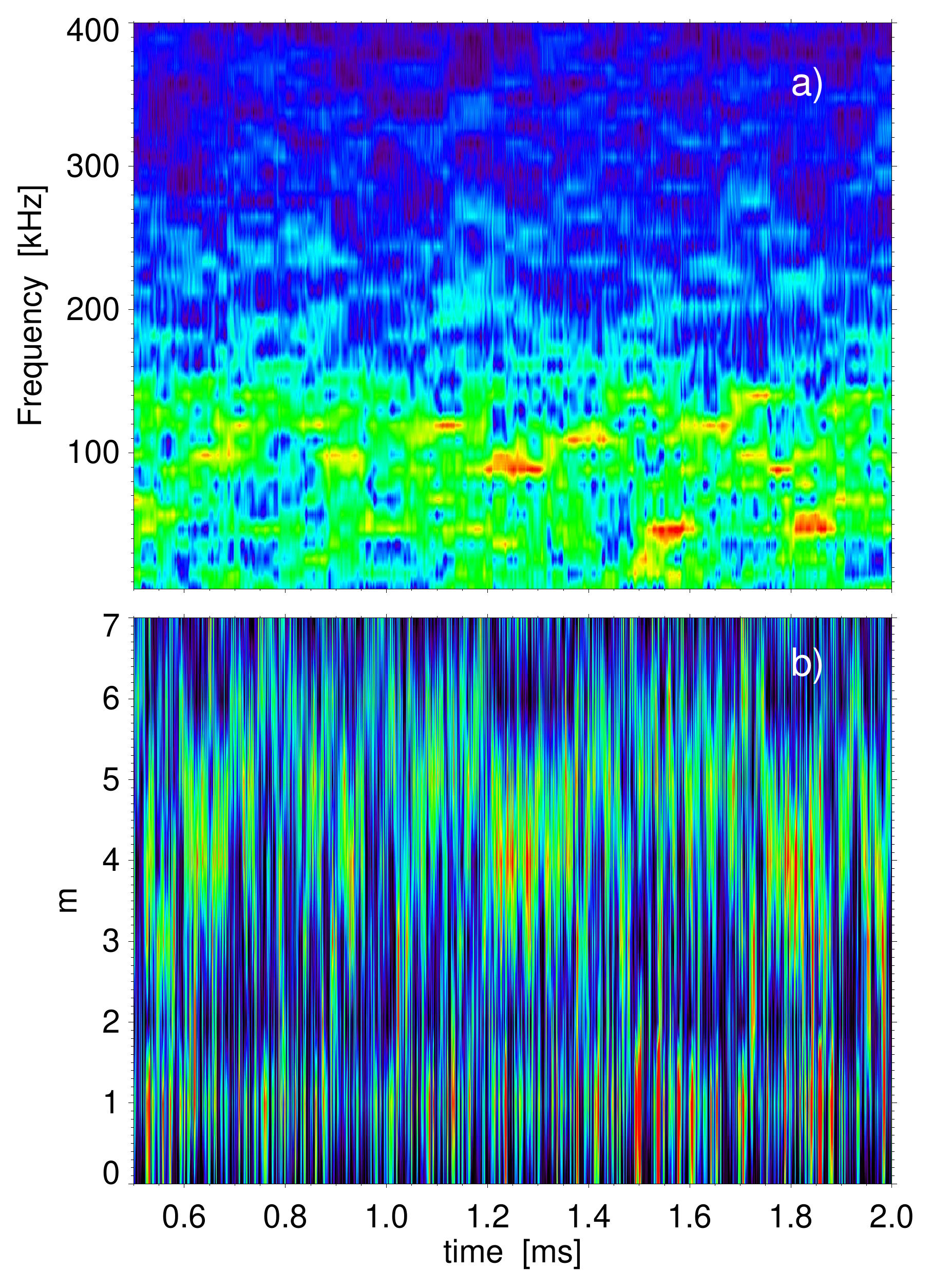}
\caption{Color-coded contours of a) spectrogram of a single probe fluctuation and
 b) time behavior of the modes amplitude.} \label{Magnetron_am_in_time}
\end{figure}

A deeper analysis of the time behavior of the fluctuations reveals that the dynamics
of the propagating waves is rather complex, as can be seen in Fig. \ref{Magnetron_am_in_time}
where the spectrogram of a single probe signal, i.e. the frequency power spectrum resolved in time,
shows that the energy related the various peaks is not constant.
In particular, the energy of the fluctuation is found to jump over the various mode numbers,
whose time resolved amplitude is shown in the bottom panel of Fig. \ref{Magnetron_am_in_time}.

No regularity can be
observed in this process, which suggests that some form of chaotic behavior
might be present, as it happens in the case of coupled oscillators.
Indeed,chaos induced by electrostatic waves in plasmas has been studied in the recent past \cite{Doveil}, and chaos control techniques were proposed and successfully tested in travelling tubes, by building barriers in phase space with small perturbations \cite{Chandre}.
A similar situation is also observed in a study of the transition
from stability to drift wave turbulence in a magnetized, low temperature,
cylindrical plasma, where a state with two broad peaks in the Fourier space is
characterized by a spatiotemporal pattern exhibiting temporally localized
defects \cite{Klinger}.

As said above, the scheme attempted in these experiments is to obtain an active
control of the plasma fluctuations in order to induce a more regular fluctuation
wavefront by creating in the 8 drivers a travelling
pattern with a defined azimuthal periodicity at a chosen frequency. In
principle, by this approach, a spatio-temporal voltage pattern which is resonant
with the natural propagating modes, can be applied. In particular, the adopted
strategy is to drive one single mode with a local electrode system (in contrast with the generally
used global systems \cite{Klinger}) to control the whole
plasma behavior.

\begin{figure}
\centering
\includegraphics[width=1.\columnwidth]{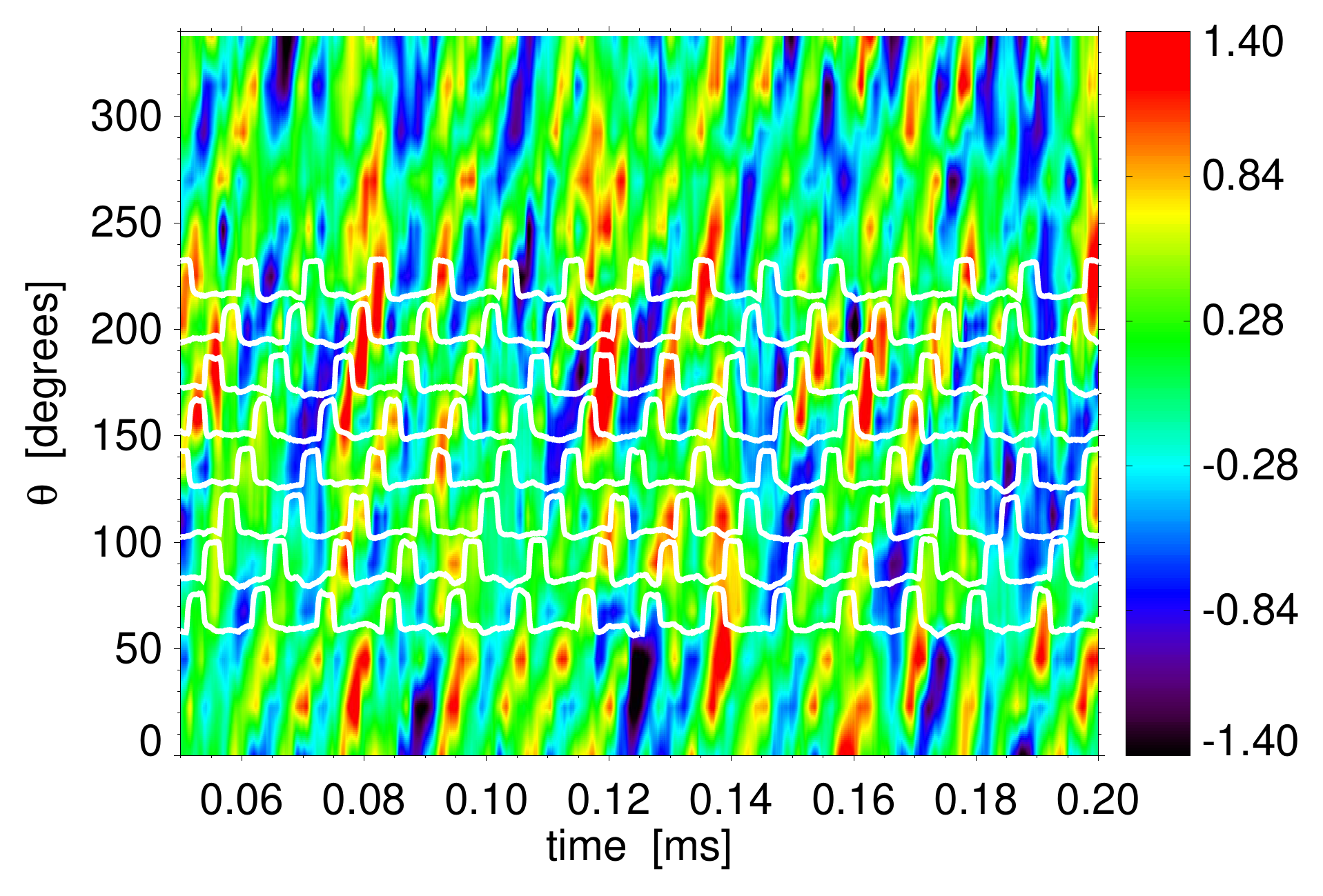}
\caption{Contour plot of the time trace of the 16 signals (color scale is in Volts).
The overplotted white curves are the voltage signals of the 8 electrodes.
A
propagating pattern with $m=4$ periodicity can be observed both on the probes and the electrodes.}
\label{Magnetron_drivers_signals}
\end{figure}

In particular, we decided to operate on the $m=4$ and $m=5$ modes at different frequencies. An example of the
pattern of pulses applied to the drivers can be seen in Fig.
\ref{Magnetron_drivers_signals} overplotted (in false units) on the color-coded contour of the probe signals.

\begin{figure}
\centering
\includegraphics[width=1.\columnwidth]{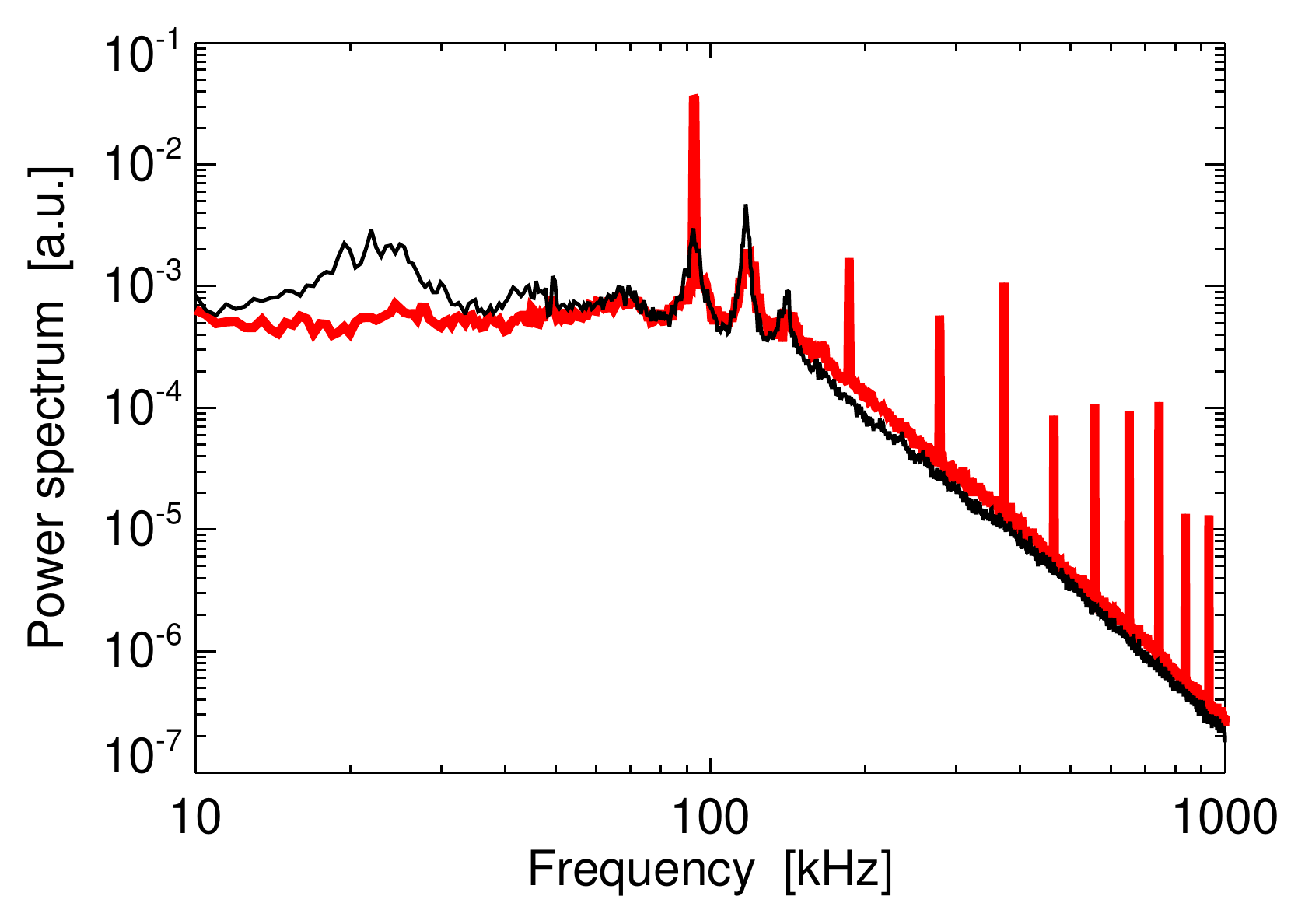}
\caption{Power spectrum of a single probe signal for a reference experimental condition (black),
i.e. with no current driven on the electrodes,  and with
an applied $m=4$ mode at 93 kHz (red).
and} \label{Sf_confronto}
\end{figure}

\begin{figure}
\centering
\includegraphics[width=1.\columnwidth]{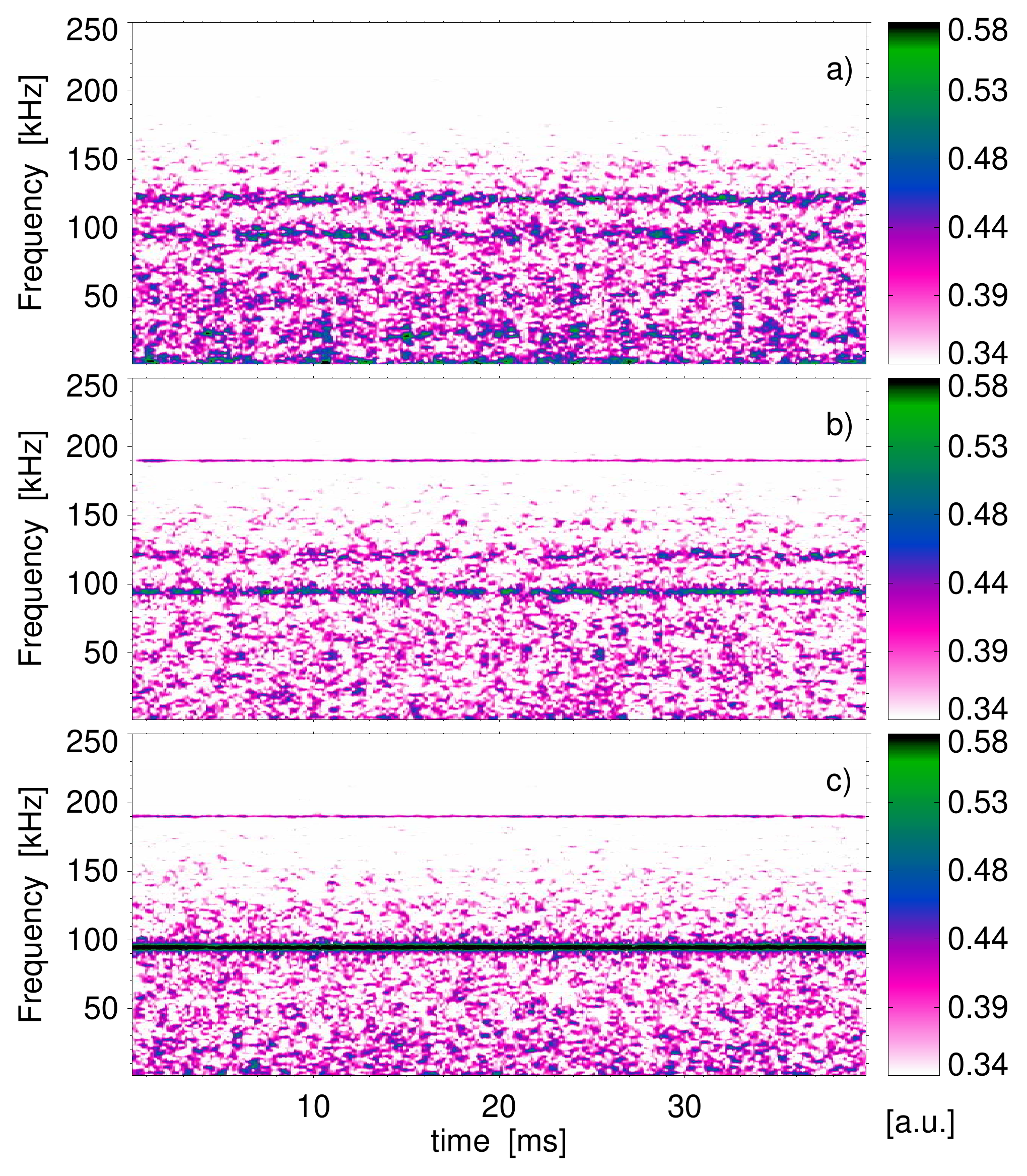}
\caption{Color-coded contours of the spectrogram (i.e. power spectrum $vs$ time) of a single
$V_f$ signal with: a) no external perturbation, b) with a counter rotating $m=4$ mode and c)
with a co-rotating $m=4$ mode.
and} \label{spectr_co_controtante}
\end{figure}

An example  of what happens to the spectrum when a resonant travelling $m=4$ mode
is applied is shown in Fig. \ref{Sf_confronto}. The signal
 is chosen to be on the opposite half side of the probe array with
respect to the half array where also the electrodes lay (see Fig. \ref{Magnetron_probes_electrodes}).
The spectrum exhibits a sharp peak at the applied frequency and at its higher harmonics,
along with a significant reduction of the power associated to the other modes, in particular to that
of $m=0$ mode at around 20 kHz, which appears almost fully depressed.
Negligible effect on the high frequency broadband part of the
spectrum is observed, differently from what
was found in Ref. \cite{Schroder}
where a preselected mode was enhanced to the expense of broadband low-frequency
spectral components.
It is important  to say that these results are obtained only with the
application of a perturbation which is co-rotating with the natural
fluctuations, while a counter-rotating perturbation gives almost no effect. This is
shown in Fig. \ref{spectr_co_controtante},
where the spectrogram of the $V_f$ fluctuation
is presented for the cases with no external and with a co-rotating and a counter-rotating $m=4$ mode at 95 kHz.
Only in Fig. \ref{spectr_co_controtante}(c) (i.e., in the co-rotating case)
the induced peak is found to largely affect the full fluctuation spectrum.

\begin{figure}
\centering
\includegraphics[width=1.\columnwidth]{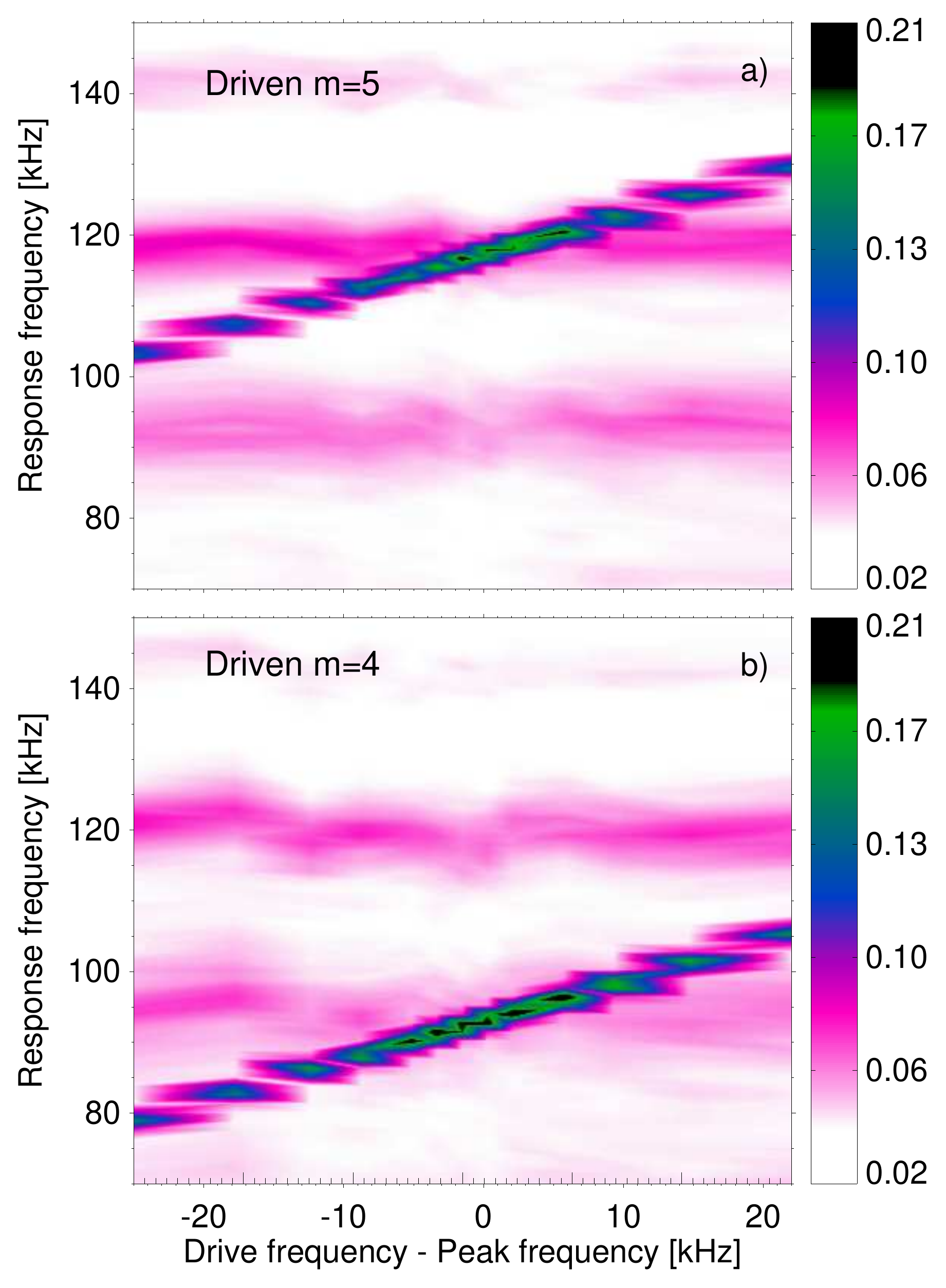}
\caption{Power spectra of a single floating potential probe as a function of
the driving frequency for an applied $m=4$ (a) and $m=5$ (b) mode.} \label{Magnetron_Sfdrive_Sfresponse}
\end{figure}

Actually the behavior of the peak associated to the induced perturbation is a function of the applied frequency itself
and of the corresponding wavelength of the rotating pattern on the
drivers.
In Fig. \ref{Magnetron_Sfdrive_Sfresponse} the power spectrum of the single floating
probe signal is plotted in a color coded
contour as a function of the driving frequency, relative to the natural one for the considered
mode. It can be observed that the height of the
additional sharp peak is a function of the driving frequency;
the peak has, indeed, a maximum when the driving frequency equals that of the
naturally occurring $m=4,5$ mode, i.e when the perturbation is resonant with the
plasma.

\begin{figure}
\centering
\includegraphics[width=1.\columnwidth]{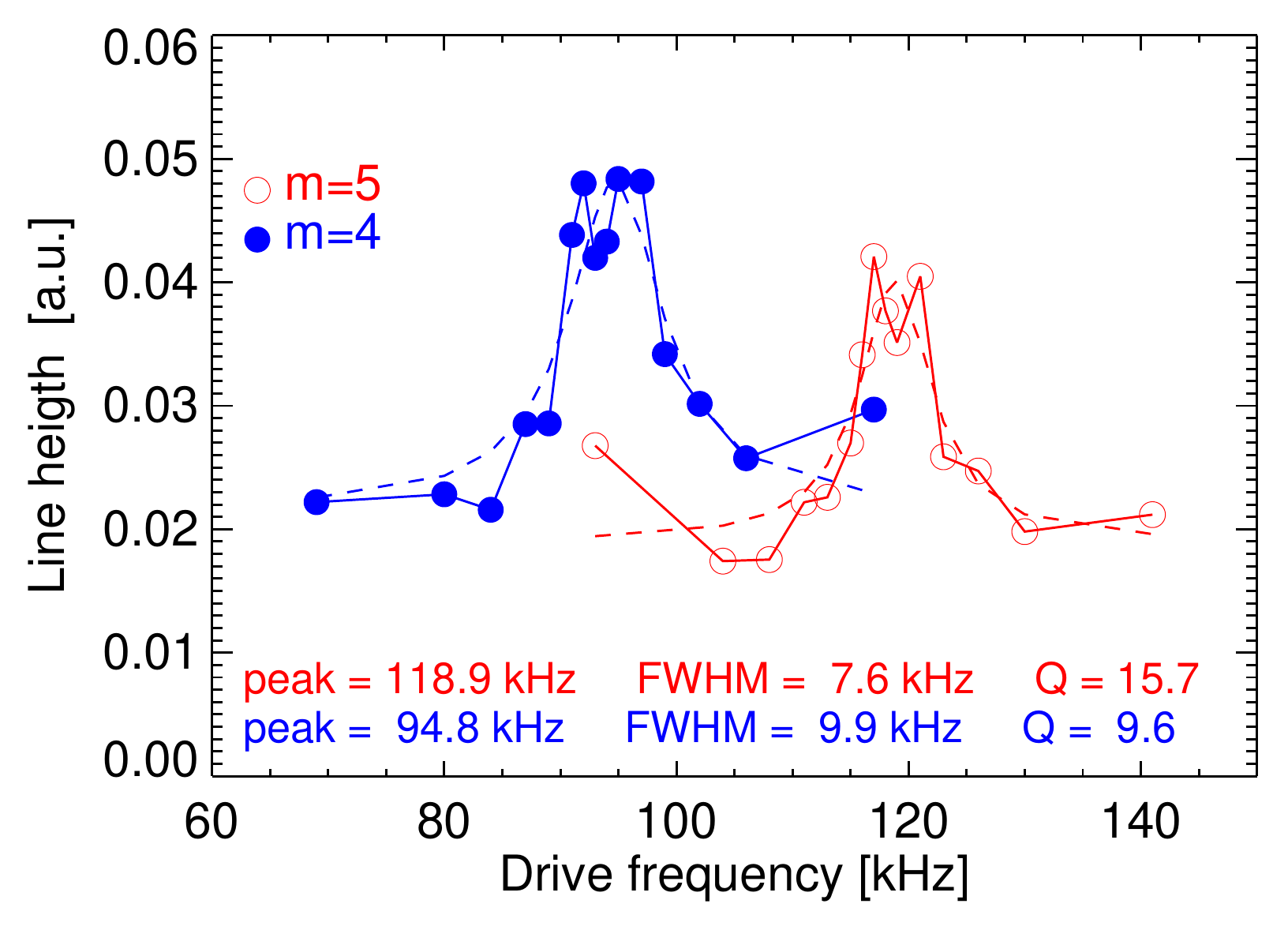}
\caption{The eight of the additional peak plotted as a function of the driving
frequency for the $m=4$ (full dots) and $m=5$ (open circles) mode. Dashed lines represent
the Lorentzian interpolation,
whose properties (Q factor, FWHM and peak frequency) are also indicated.}
\label{Magnetron_maximum_peak}
\end{figure}

If the height of the additional peak is plotted as a function of the
driving frequency, as is shown in Fig. \ref{Magnetron_maximum_peak}, a
bell-shaped curve is indeed obtained, which is reminiscent of the response of a
resonator. The ratio of the curve width to the peak frequency yields a quality
factor $Q=\omega_0/\Delta\omega\approx 10$, depending on the mode number considered.

\begin{figure}
\centering
\includegraphics[width=1.\columnwidth]{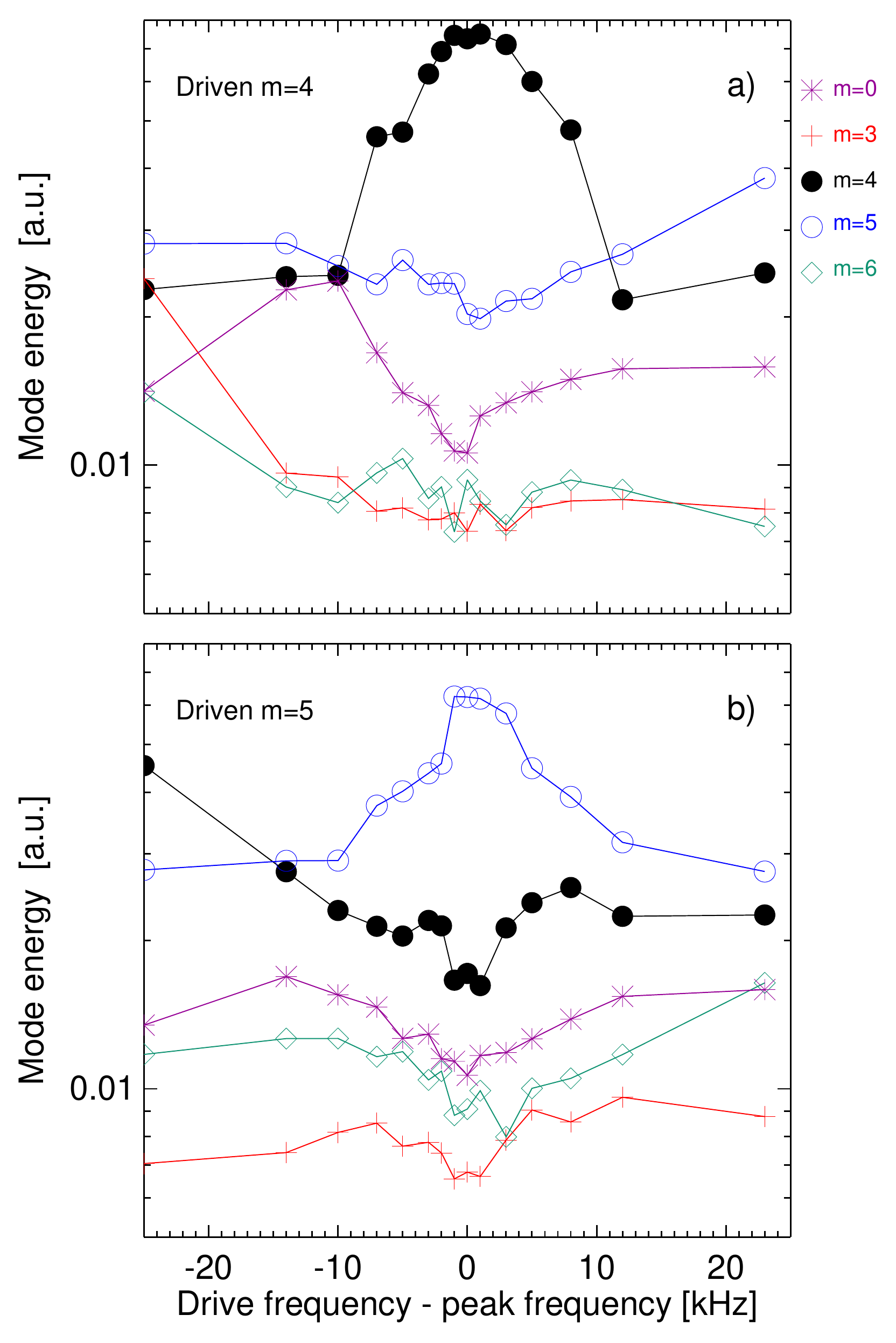}
\caption{Energy of the $m=0,3,4,5,6$ modes
plotted as a function of the frequency difference between drive and
spontaneous fluctuation: a) and b) refer to cases with a driven $m=4$ and $m=5$ mode, respectively.}
\label{Magnetron_modes_amplitude_vs_fr}
\end{figure}

Turning to the peak in the spectrum associated to the drift-wave mode, it is found that also its energy is
strongly dependent on the driven frequency. This is shown in Fig. \ref{Magnetron_modes_amplitude_vs_fr}, where
the amplitude of the main $m=0,3,4,5,6$ modes is shown for various applied frequencies to the driven electrodes (what is
actually used is the difference with respect to the natural one), and
with a $m=4$ and a $m=5$ applied pattern.
For both cases, when the natural frequency is matched, i.e. when a resonant condition is achieved,
a large enhancement of the mode is found, with part of the energy being taken from the other ones.
The system thus seems to be pushed towards a more regular state.
As previously deduced from the analysis of the simple $S(f)$ spectrum,
a strong effect is confirmed to be found on the $m=0$ mode, whose amplitude is reduced by a factor
two, but also the others not-driven modes exhibit a sensible reduction.

\begin{figure}
\centering
\includegraphics[width=1.\columnwidth]{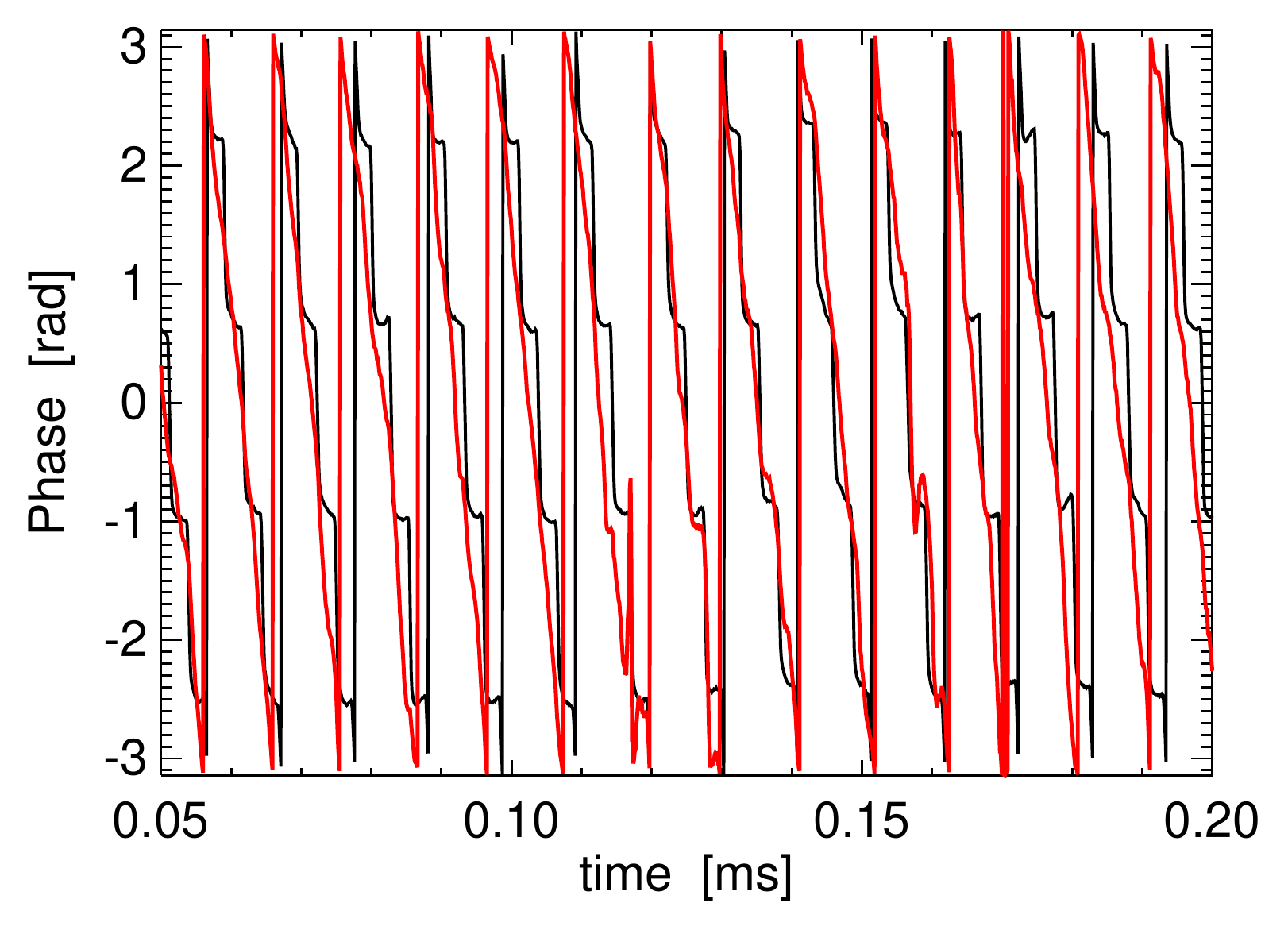}
\caption{Time evolution of the phase of the $m=4$ drive (black) and $m=4$ response (red) in resonant
condition.} \label{Magnetron_Phase_drive&response}
\end{figure}

The effect of the perturbation on the plasma, in the resonant condition is displayed in Fig.
\ref{Magnetron_Phase_drive&response}, which shows the phase $\alpha_{drive}(t)$
of the $m=4$ mode applied to the drivers (in black) and the phase of the
$\alpha_{m}(t)$ of the $m=4$ mode as measured by the sensors.
It is possible to see, qualitatively, that the drive and the response are
synchronized only for a fraction of the time.

\begin{figure}
\centering
\includegraphics[width=1.\columnwidth]{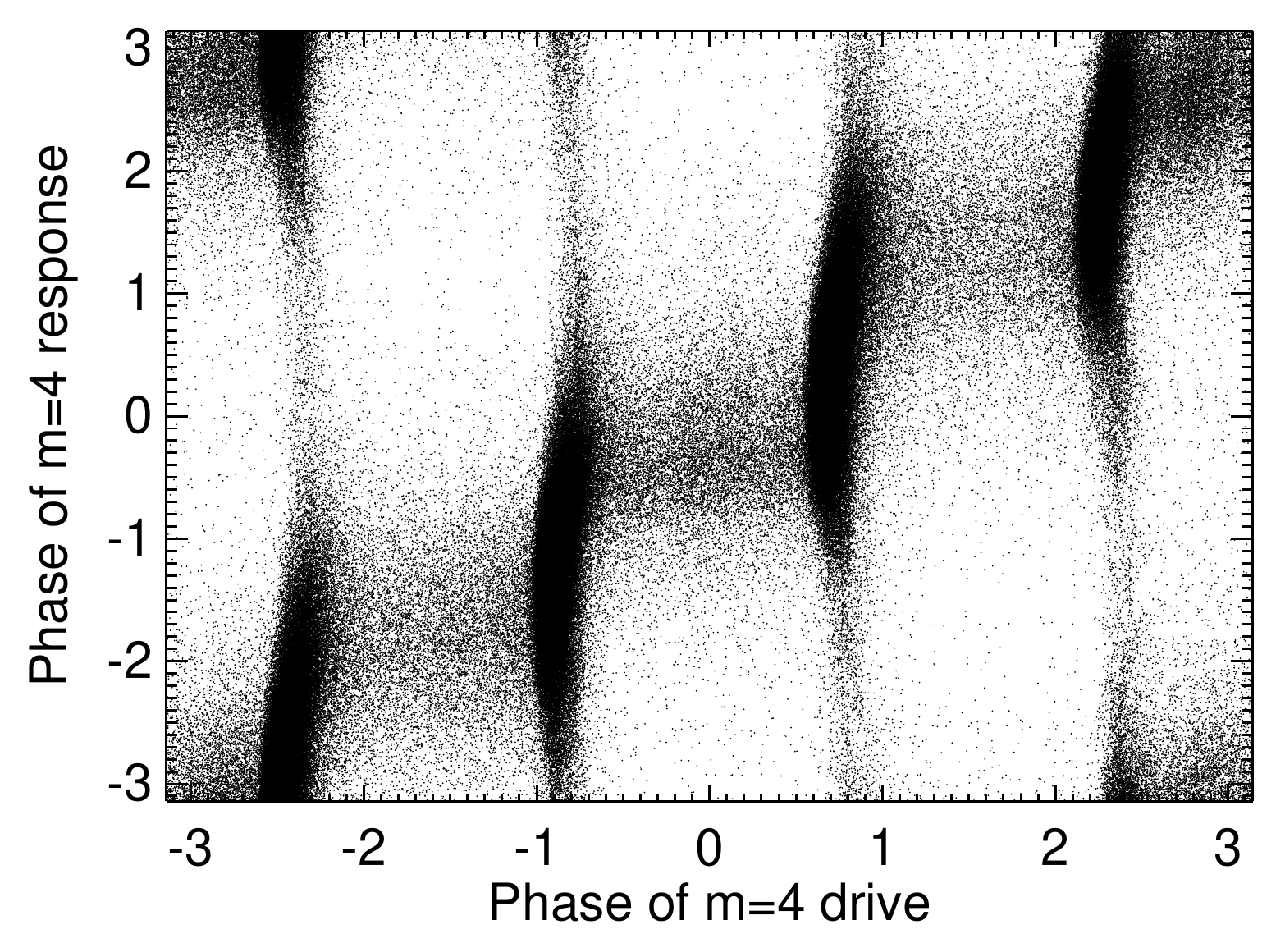}
\caption{Phase of the $m=4$ drive vs. $m=4$ response.}
\label{Magnetron_Phdrive_vs_Phresp}
\end{figure}

\begin{figure}
\centering
\includegraphics[width=1.\columnwidth]{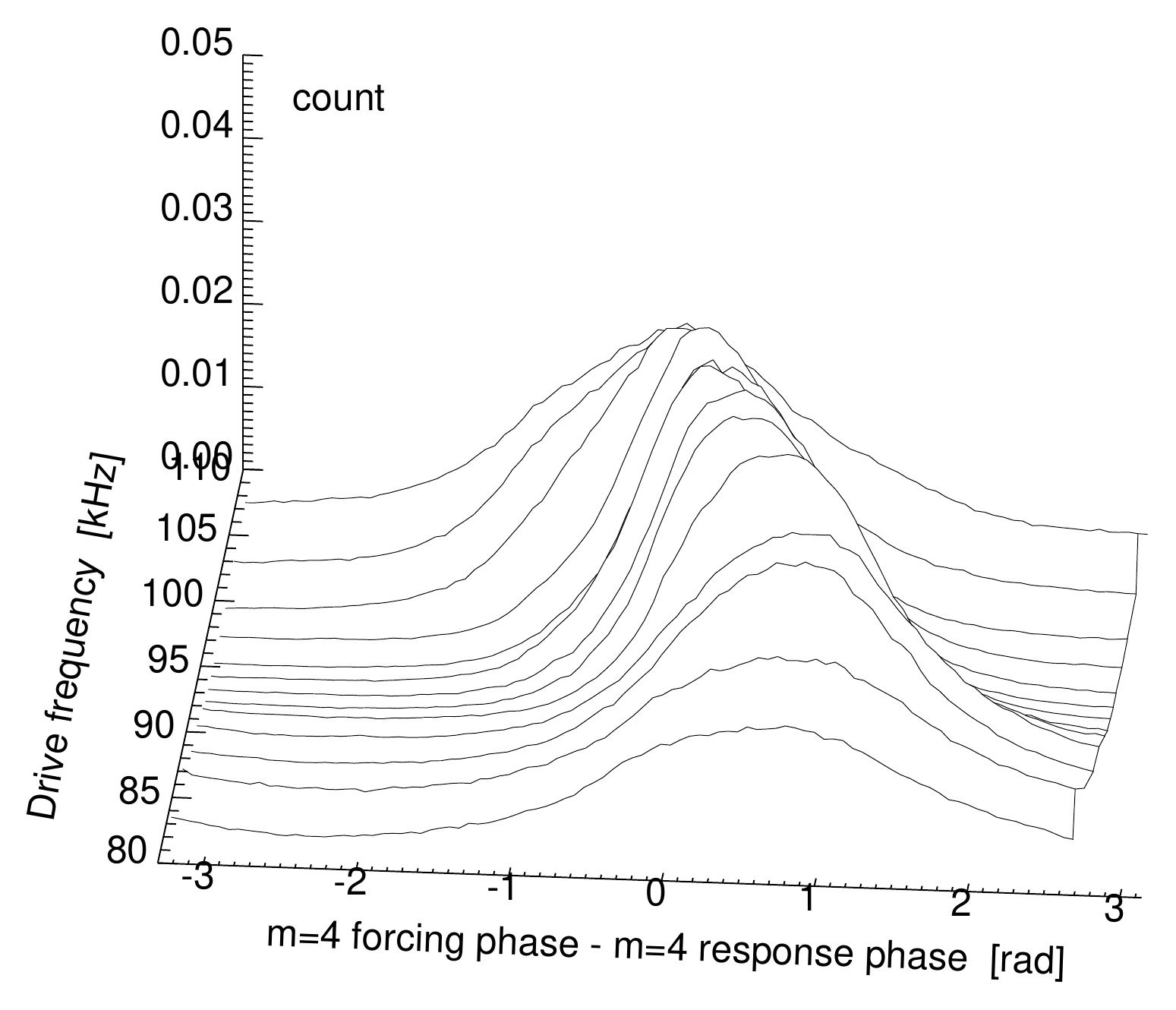}
\caption{Distribution of phase difference between the $m=4$ drive and the $m=4$
response for different values of the driving frequency.}
\label{Magnetron_phases_distribution}
\end{figure}

This is more clearly seen in Fig. \ref{Magnetron_Phdrive_vs_Phresp}, where the
phase of the response is plotted versus the phase of the drive for the whole
time record.
The black vertical stripes are due to the fact that the drive phase does not
vary linearly, since the drive is not sinusoidal. Apart from this effect, it is
also possible to see a diagonal darker band, which shows that there exists a
preferred phase relationship between drive and response. Such relationship
however occurs only intermittently, so that some points are present also in
other regions of the plot.
Indeed, the occurrence of the phase synchronization, as defined by the presence of a
preferred value of the drive-response phase difference, is  found to depend on
the drive frequency. Indeed, the phase difference distribution  is more peaked
when the drive frequency matches that of the naturally occurring mode (i.e. in a resonant condition),
with a peak value around $\pi /10$, confirming the
existence of a partial phase synchronization, and
becomes more and more flat as the frequency is moved away from the resonance
condition. This can be observed in Fig. \ref{Magnetron_phases_distribution},
where the phase difference distribution is plotted for different driving
frequency values.

\begin{figure}
\centering
\includegraphics[width=1.\columnwidth]{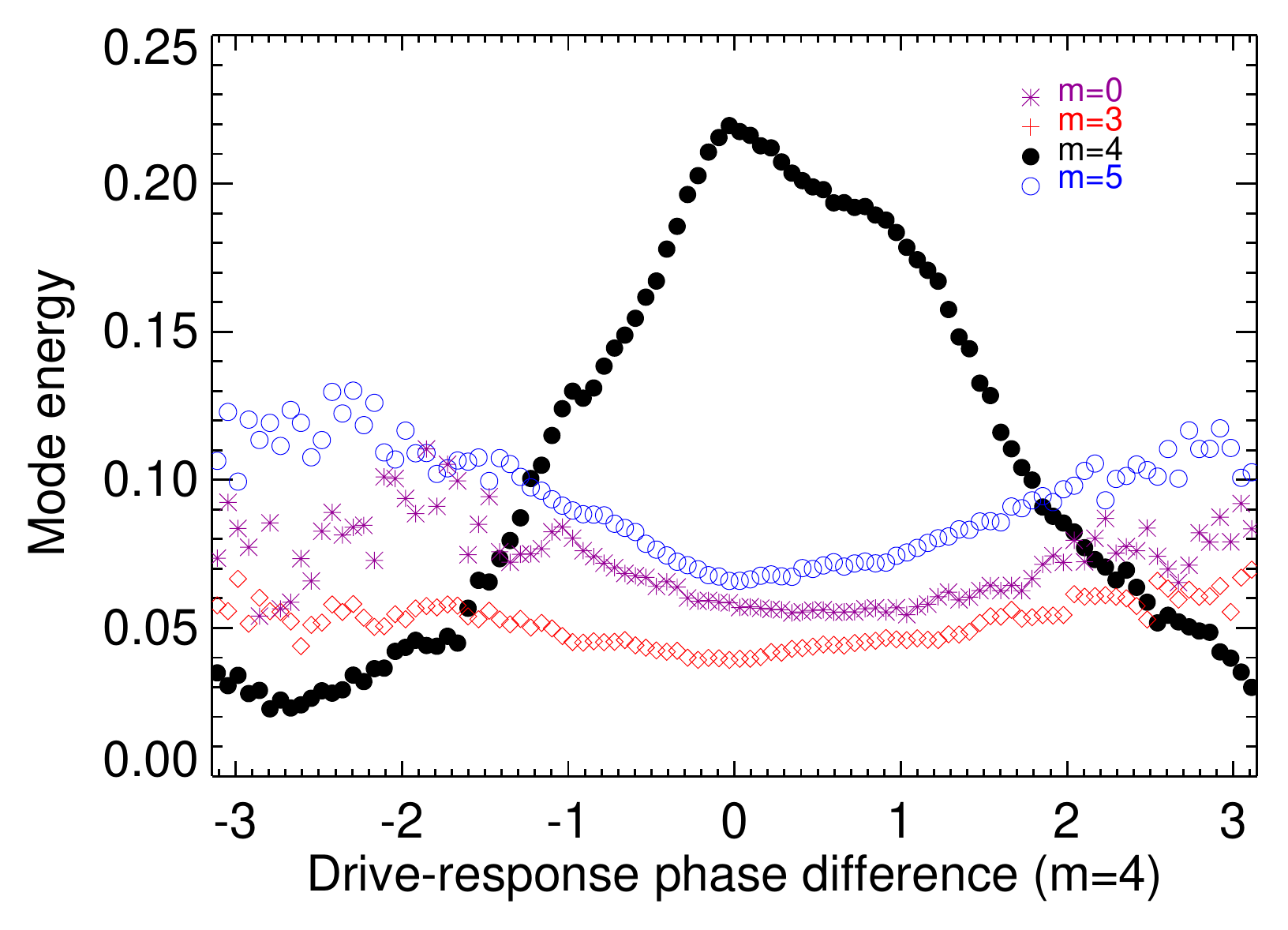}
\caption{Modes amplitude as a function of the phase difference between the $m=4$ drive and the $m=4$
response in resonant condition.}
\label{Magnetron_ampl_vs_phasediff}
\end{figure}

Finally, Fig. \ref{Magnetron_ampl_vs_phasediff} shows the relative mode
energy (i.e. the mode energy normalized to the total energy of the modes) for the case of an applied
$m = 4$ mode in a resonant condition, along with that of the $m =0,3,5$ modes, plotted as a function of the
drive-response phase difference between the spontaneous and the induced fluctuations.
It is possible to observe that when phase synchronization takes place, i.e.
when the phase difference has a value around $\pi /10$, the relative amplitude of the
driven mode ($m = 4$) is enhanced, while that of the other modes is
reduced. Such a result is reminiscent of what found in Ref. \cite{Klinger},
although it is not as strong (probably due to the drive weakness).

\begin{figure}
\centering
\includegraphics[width=1.\columnwidth]{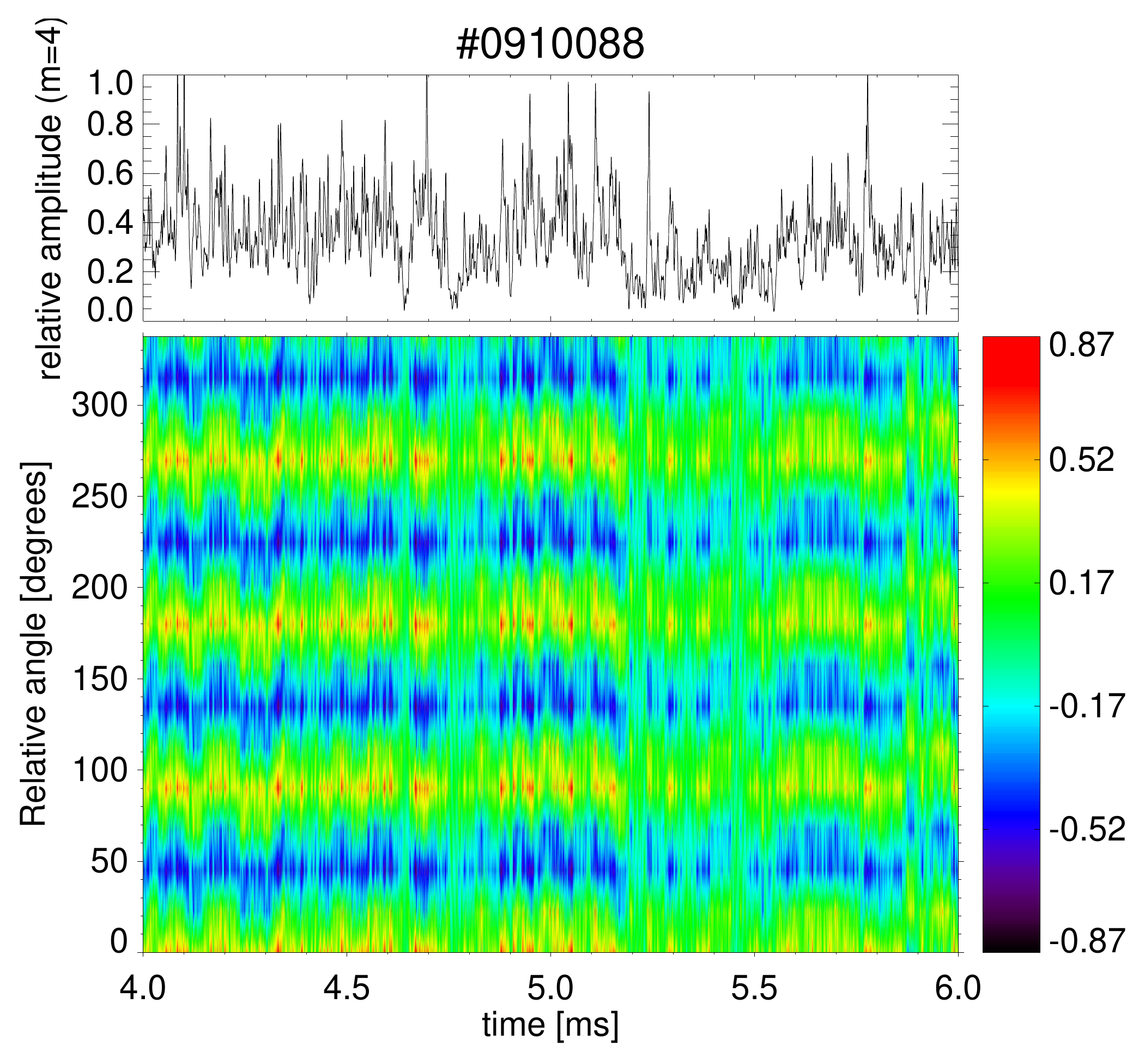}
\caption{$m=4$ mode in the reference frame of the applied perturbation
(bottom), and relative amplitude of it (top).}
\label{Magnetron_m4_in_ref_frame}
\end{figure}

In order to further establish the existence of a partial phase synchronization,
and its effect on the mode amplitude, we show in Fig.
\ref{Magnetron_m4_in_ref_frame} a color-coded plot of the quantity
\begin{equation}
x(\theta,t) = A_m(t)� cos[m\theta+\alpha_{m}(t)- \alpha_{drive}(t)]
\label{}
\end{equation}
for $m = 4$, where
$A_m(t)$ is the mode amplitude.
This quantity is the $m = 4$ mode as seen in a reference frame moving with the
applied voltage pattern \cite{Block}. The plot shows that a phase locking
between drive and response is seen only in some time intervals, where a
stationary pattern appears, and is lost in others, where the mode is found to
move with respect to the applied perturbation. The top frame, showing the
relative amplitude of the $m = 4$ mode, confirms that it increases during the
phase synchronization intervals.

\section{Conclusion}

The results of an experimental campaign aimed at controlling unstable
electrostatic
drift-waves in a low temperature magnetized plasma have been presented.
Several mode numbers are observed on the spectrum of  the floating potential fluctuation,
which are not simultaneously present all the time, but tend to act as
coupled oscillators, transferring energy from one to the other in an apparently
chaotic fashion.
By means of a local system of electrodes operating in electron current regime a
wavelike pattern propagating has been induced in resonant condition (same phase
velocity and frequency) with the naturally present modes. It has been observed
that it is possible to increase the energy of one preferred mode, reducing the
others, only if a good synchronization is obtained between the applied
perturbation and the natural one. No effects on the broadband part of the
spectrum can be seen, probably because the current collected by the electrodes
is too small if compared to the total plasma current.

We plan to apply in the next future chaos control methods based on temporal (or
spatiotemporal) feedback concepts, thus acting in closed-loop fashion in both
space and time, to obtain, from an  energetic point of view, less expensive
effects on the plasma. This approach is intended as a test bed application of
chaos and turbulence control techniques at the edge of thermonuclear fusion
devices, where electrostatic fluctuations are known to be responsible for
anomalous particles transport, in order to improve confinement properties.

\section{Appendix A}

In this appendix, the deduction of the linear dispersion relation \ref{Magnetron_dispersion_relation} will be briefly reminded.
The first important point to consider is the low
degree of
ionization, of the order of $10^{-4}$, of the plasma under study, where the collisions
with neutral atoms dominate. The main parameters of the plasma under study are listed in Table \ref{Magnetron_table},
where the collision frequencies have been evaluated by means of the expressions
for their derivation reported in Ref. \cite{Lieberman}.
\begin{table}
\centering
\begin{tabular}{lr}
\hline
\hline
Cathode radius                      \ \ \ & \ \  5 cm \\
Magnetic field  ($B$)                 \ \ \ & \ \   14.2 mT \\
Operating pressure ($P$)              \ \ \ & \ \   1  Pa \\
Plasma current ($I$)                  \ \ \ & \ \   0.6  A \\
Electron temperature ($T_e$)        \ \ \ & \ \   2  eV \\
Ion temperature ($T_i$)             \ \ \ & \ \    0.025  eV \\
Electron/ion density ($n$)          \ \ \ & \ \   $2\times 10^{17}$  m$^{-3}$ \\
Ion cyclotron frequency  ($f_{ci}$)   \ \ \ & \ \   5.4  kHz \\
Ion collision frequency ($\nu_i$)   \ \ \ & \ \   59  MHz \\
Electron cyclotron frequency ($f_{ce}$)   \ \ \ & \ \   400  MHz \\
Electron collision frequency ($\nu_{e}$)   \ \ \ & \ \   190  MHz \\
\hline\hline
\end{tabular}
\caption{Main plasma parameters in the magnetron sputtering device.}
\label{Magnetron_table}
\end{table}

An important feature of this plasma is that only the electrons can be considered as magnetized.
Indeed, given the measured plasma parameters, the electron and ion Larmor radius result
$r_{Le}\sim0.5$ mm and $r_{Li}\sim10$ mm, respectively (the ion temperature in this
non-thermal plasma can be assumed equal to that of the neutral gas, i.e. about 0.025 eV).
These values must be compared to the mean free paths for electrons and ions $\lambda_{e}$, $\lambda_{i}$ which are 20 and 90 mm,
respectively \cite{Lieberman}.
As a consequence, the Hall parameter results $\gg$ 1 for electrons and  around 1 for ions in the magnetic trap volume.
Therefore, ions are unmagnetized due to a Larmor radius comparable with their
mean free path, whereas electrons are magnetized and are subjected to an $E \times B$ drift.
The drifting electrons give rise to an azimuthal Hall current $I_{E\times B}$, roughly estimated as:
$I_{E \times B}= e n_e  \pi a^2 E/B \sim 0.2A$, with $a$ the radius of the equivalent circular cross-section
for the azimuthal current.
The corresponding magnetic field evaluated at $r=a$ is: $B=\mu_0 I_{e \times B}/(2\pi a)\sim 4 \mu T$,
that is negligible compared with the values measured in the magnetic trap (Fig. \ref{schema_magnetron}).

In two previous papers \cite{Martines01,Martines04} we interpreted
the observed electrostatic fluctuations in terms of an $\textbf{E}
\times\textbf{B}/$density gradient instability, also known as \textit{neutral
drag} instability. A linear dispersion relation was obtained for a low  $\beta_e$, weakly ionized plasma,
in a slab geometry with a straight magnetic field $B$ along the $z$ axis, the $x$ axis being in
the direction of the density and potential gradients. In such geometry the $y$ direction corresponds
to the azimuthal direction.
The continuity equation for ions and
electrons is considered:

\begin{equation}
\frac{\partial n}{\partial t}+ \nabla \cdot(n \textbf{v}_{e,i})=Zn
\label{Magnetron_continuity_eq}
\end{equation}

along with their equations of motion, where
electron inertia and the ion temperature $T_i$ are neglected:

\begin{equation}
\frac{\partial \textbf{v}_i}{\partial t}+(\textbf{v}\cdot\nabla )\textbf{v}=
\frac{e}{M}(\textbf{E}+\textbf{v}\times\textbf{B})-\nu_i\textbf{v}
\label{Magnetron_motion_i_eq}
\end{equation}

\begin{equation}
0=-\frac{T}{nm} \nabla
n-\frac{e}{m}(\textbf{E}+\textbf{v}\times\textbf{B})-\nu_e\textbf{v}
\label{Magnetron_motion_e_eq}
\end{equation}

The right-hand side term of eq. \ref{Magnetron_continuity_eq} describe
ionization events, with $Z$ indicating the ionization rate. The same symbol $n$
is used for electron and for ion densities (i.e. $n_e=n_i=n$), because quasineutrality is assumed
due to the low frequency of the waves under consideration
($\omega/\omega_p<<1$, where $\omega_p$ is the electron plasma frequency). The
neutral gas dynamics is assumed not to be affected by collisions of electrons and ions with neutrals.

By a standard procedure, which includes linearization, the dispersion relation:
\begin{equation}\begin{split}
\omega^2 +
\omega[\frac{\Omega_i}{\nu_i}(k_x-\frac{\Omega_i}{\nu_i}k_y)v_E+i(\nu_i-Z)]-
\\
-k^2C_s^2+Z\nu_i-i\Omega_i[(\frac{\Omega_i}{\nu_i}k_y-k_x)v_E+k_xv_{de}]=0
\end{split}\end{equation}
is thus determined (here $C_s$ is the ion sound speed, $C_s\simeq(T_e/m)^{1/2}$, $v_E$ is the
$\textbf{E}\times\textbf{B}$ velocity, $\nu_i$ is the collision frequency of
ions with neutrals, $\Omega_i$ is the ion cyclotron frequency multiplied by
$2\pi$, k is the wavevector and $v_{de}$ is the electron diamagnetic drift
speed $v_{de}=-T_e/(enB)(dn/dx)$).

\

\end{document}